\documentclass [ prl, twocolumn, amsmath, showkeys, nofootinbib, floatfix ] {revtex4}

\usepackage{graphicx}
\usepackage{xfrac}
 \usepackage{mathptmx}      
\usepackage{amsmath}
\usepackage{array}

\newcolumntype{W}{>{\centering\arraybackslash}m{2.4cm}}
\newcolumntype{C}{>{\centering\arraybackslash}m{2.0cm}}

\newcommand{\degree}{\ensuremath{^\circ}}
\begin{document}

\title{Avalanches on a Conical Bead Pile: Scaling with Tuning Parameters}

\author{S.Y. Lehman}
\author{Elizabeth Baker}
\author{Howard A. Henry}
\author{Andrew J. Kindschuh}
\author{Larry~C. Markley}
\author{Megan B. Browning}
\author{Mary E. Mills}
\author{R. Michael Winters IV}
\author{D.~T.~Jacobs}

\affiliation{Physics Department, The College of Wooster, Wooster OH 44691 USA}

\begin{abstract}
Uniform spherical beads were used to explore the scaling behavior of a granular system near its critical angle of repose on a conical bead pile. We found two tuning parameters that could take the system to a critical point.  The existence of those tuning parameters violates the fundamental assumption of self-organized criticality, which proposed that complex dynamical systems self-organize to a critical point without need for tuning. Our avalanche size distributions were well described by a simple power-law with the power $\tau = 1.5$ when dropping beads slowly onto the apex of a bead pile from a small height.  However, we could also move the system from the critical point using either of two tuning parameters:  the height from which the beads fell onto the top of the pile or the region over which the beads struck the pile.  As the drop height increased, the system did not reach the critical point yet the resulting distributions were independent of the bead mass, coefficient of friction, or coefficient of restitution. All our apex-dropping distributions for any type of bead (glass, stainless steel, zirconium) showed universality by scaling onto a common curve with $\tau = 1.5$ and $\sigma = 1.0$, where $1/\sigma$ is the power of the tuning parameter. From independent calculations using the moments of the distribution, we find values for $\tau = 1.6 \pm 0.1$ and $\sigma = 0.91 \pm 0.15$.  When beads were dropped across the surface of the pile instead of solely on the apex, then the system also moved from the critical point and again the avalanche size distributions fell on a common curve when scaled similarly using the same values of $\tau$ and $\sigma$.  We also observed that an hcp structure on the base of the pile caused an emergent structure in the pile that had six faces with some fcc or hcp structure. 

\keywords{bead pile; avalanches; scaling behavior; granular material; universality; critical behavior; critical exponents; self-organized criticality}
\end{abstract}

\maketitle

\section{Introduction}
	\label{intro}

Some have proposed that granular (and other) systems self-organize to true critical points \cite{BakHowNatureWorks,BakPRA} like dry sand forming a conical pile at its critical angle of repose; others have suggested that some tuning parameters must exist in order to bring a granular system to the critical point \cite{JaegerScience}.  In this paper, we present two such tuning parameters and the remarkable scaling behavior associated with them. Critical points are characterized by a simple power-law behavior for relevant properties when brought sufficiently near the critical point by varying some tuning parameter like temperature or concentration.  In many systems (liquids, ferromagnets, polymer solutions, etc.), a wide range of scaling behavior occurs as characterized by Widom \cite{WidomJPC} and later formalized by renormalization group theory \cite{FisherRMP}.   The particulars of the system determine where the critical point is located, but when sufficiently near the critical point then the system's properties (like correlation length or susceptibility) scale with temperature as a power-law with powers that are universal values for all types of systems \cite{GreerARPC,KumarPhysRep}.

A granular system behaves in some ways like a liquid with an ability to flow and in some ways like a solid with a stable fixed structure if undisturbed.  Bak, Tang, and Wiesenfeld \cite{BakHowNatureWorks,BakPRA} proposed that granular piles would self-organize unaided by any tuning parameter to a critical point and would avalanche with size and temporal distributions that would be described by simple power-laws. As grains are added to a pile, a much larger percentage of avalanches would be small rather than large and the avalanche size distribution would follow a power-law relation \cite{LauritsenPRE,ZapperiPRL}
\begin{equation}
	P(s)=P_o s^{-\tau},	\label{powerlaw}
\end{equation}
where $P(s)$ is the fraction of avalanches of size $s$,  $P_o$ is a constant, and $\tau$ is the power.  The power   is predicted by mean-field theory (MFT) to be 3/2 when the dimension of the system is greater than the critical dimension of four \cite{LeDoussalPRE} but to be 1.33 in three dimensions \cite{LubeckPRE}.  This theory of self-organized criticality (SOC) has been applied to many systems including earthquakes \cite{BakPRL,GutenbergAnnDiGeof}, neuronal avalanches \cite{PajevicPLOS}, and stock market trends \cite{GabaixNature}.  

The conditions under which a system exhibits SOC and simple power law distributions for the size and duration of its response have been the subject of numerous theoretical and numerical simulations with relatively few experimental investigations \cite{AegerterPRE,GerashchenkoJSM,RamosPRL}; a history we have described previously in detail \cite{CostelloPRE}.  

Here we have undertaken a systematic investigation of a simple experimental system that models a sandpile---a pile of spherical beads---in order to explore when the system can be described by a pure power-law.  We have experimentally found two parameters that tune the system away from the critical point. The existence of these tuning parameters that bring a system to its critical point violates a fundamental assumption of SOC.

In the experiment, we formed a conical bead pile on a circular base, then dropped one bead at a time onto the pile (either at the apex or across the pile) and measured the number of beads falling off the pile, which we took to be proportional to the size of the avalanche.  We varied the height at which beads were dropped, the location where they landed on the pile, the type of bead, and the underlying base geometry of the pile.
We use the analytic tools developed for equilibrium systems near a critical point and apply them to this non-equilibrium, granular system.   In addition to looking at values of the critical exponent $\tau$ as the slope of a line on a log-log plot, we also look for universal scaling functions that can collapse our data onto a common curve and provide a more compelling case for an exponent's value.  We show that these scaling functions collapse our data on several different types of beads remarkably well over several decades of avalanche size.

\section{Theory}
	\label{theory}
	
As a granular system moves from the critical point and a pure power-law, the avalanche size distribution is typically characterized by a function determined from MFT in which the system departs from a power law at a characteristic  avalanche size $s_o$   \cite{LauritsenPRE,ZapperiPRL,GhaffariPRE}, so that
\begin{equation}
	P(s)=P_o s^{-\tau} e^{-s/s_o} .	\label{powerlawRollOff}
\end{equation}
When $s_o$ is large, the function collapses to a simple power law.  We have previously experimentally verified this behavior in uniform glass beads \cite{CostelloPRE} and found a proportional relationship between $1/s_o$ and the height $h$ from which beads are dropped onto the apex of a near-critical bead pile.  To emphasize the role of $h$ as a tuning parameter, the argument of the exponential $-s/s_o$  may be rewritten as $-Ash^{1/\sigma}$ , where $A$ is a proportionality constant and $\sigma$ is a universal exponent \cite{DahmenPRL}.  More generally, the exponential function of the product $sh^{1/\sigma}$  may be replaced by an unspecified scaling function $\Re\left(sh^{1/\sigma}\right)$  for any tuning parameter $h$, so that the distribution becomes
\begin{equation}
	P(s)=P_o s^{-\tau} \Re\left(sh^{1/\sigma}\right)	\label{powerlawScaling}
\end{equation}
A random-field Ising Model has shown the same universal critical exponents and scaling functions regardless of whether a system is in equilibrium or non-equilibrium indicating that these both belong to the same universality class in 3D \cite{LiuEPL}.  Sethna et al. \cite{SethnaNature} have used a renormalization-group framework to analyze the avalanche behavior of crackling noise (including earthquakes, crumpled paper, and Barkhausen noise in magnets) and to search for the universality class governing crackling.  They emphasize that scaling functions can be more sensitive than critical exponents to discriminate between different universality classes because scaling functions contain much more information than do the small set of numbers which are the critical exponents.  This framework provides specific predictions for the moments $\langle s \rangle$  and $\langle s^2 \rangle$  of the avalanche size distribution.  For a distribution given by a power-law modified by a scaling function $\Re\left(sh^{1/\sigma}\right)$ as in Eq. (\ref{powerlawScaling}), then the moments of the distribution $\langle s \rangle$ and $\langle s^2 \rangle$, defined below, may be rewritten using the scaling function and using a change of variables in the integration to give 
\begin{subequations}	\label{moments}
	\begin{align}
		&\langle s \rangle =\int_{\text{all s}}s P(s)ds=\int_{\text{all s}}s P_o s^{-\tau} \Re\left(sh^{1/\sigma}\right) ds \propto h^{\frac{\tau-2}{\sigma}}	\label{Smom}
		\text{, and}\\
		&\langle s^2 \rangle=\int_{\text{all s}}s^2 P(s)ds=\int_{\text{all s}}s^2 P_o s^{-\tau} \Re\left(sh^{1/\sigma}\right) ds \propto h^{\frac{\tau-3}{\sigma}}	\label{S2mom}
	\end{align}
\end{subequations}
for \emph{any} scaling function $\Re$. Thus, if the tuning parameter and scaling are valid, then values for the universal exponents $\tau$ and $\sigma$ can be determined from the dependence of the moments of the distribution $\langle s \rangle$  and $\langle s^2 \rangle$ on the tuning parameter $h$.  

The scaling function also provides a means of collapsing the data obtained at different values of the tuning parameter onto a single curve.  Multiplying Eq.~(\ref{powerlawScaling}) by $h^{-\tau/\sigma}$  yields
\begin{equation}
	P(s) h^{-\tau/\sigma}=P_o \left( sh^{1/\sigma}\right)^{-\tau} \Re\left(sh^{1/\sigma}\right),	\label{collapse1}
\end{equation}
so that a plot of $P(s) h^{-\tau/\sigma}$ versus $sh^{1/\sigma}$ should collapse all the distribution data from different heights and bead types onto the same curve.  Alternatively, Eq. (\ref{powerlawScaling}) may be multiplied by $s^\tau$ instead to obtain
\begin{equation}
	P(s) s^\tau=P_o  \Re\left(sh^{1/\sigma}\right),	\label{collapse2}
\end{equation}
which will provide a collapse on a plot of $P(s) s^\tau$ versus $sh^{1/\sigma}$.  The two methods of collapse are complementary since we are interested in the values of two independent universal exponents, $\tau$ and $\sigma$. 

\section{Experimental Details}
\label{experiment}

\begin{figure}[b]
  \includegraphics{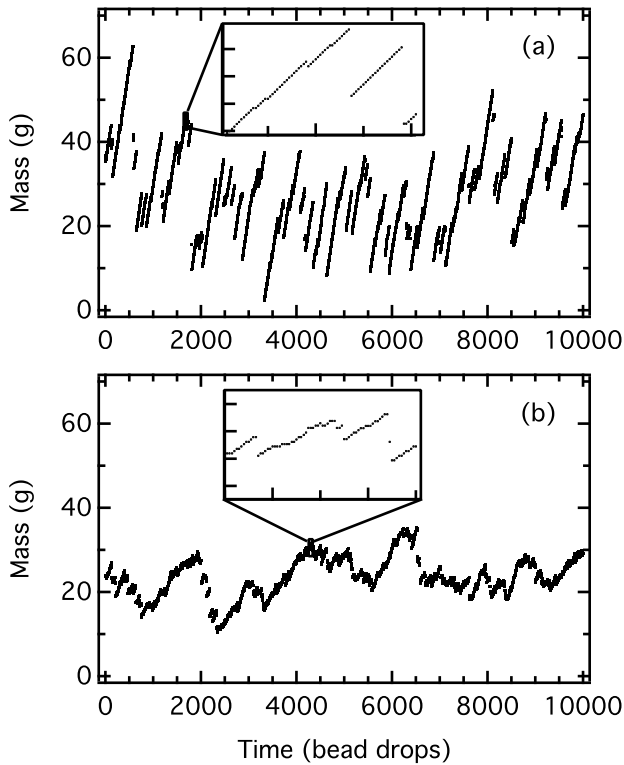}
\caption{The mass of a conical pile of zirconium beads after each bead drop.  (a) The beads are dropped from a height of 1.5 cm onto the apex.  (b) The beads are dropped from a height of 2 cm (as measured from the apex of the pile) and over a width of 7.6 cm across the surface.  Both plots assign the vertical scale's zero arbitrarily.  The inset graphs expand a narrow sliver (80 bead drops) of the main graph; insets in both (a) and (b) have a vertical full scale of 4 g }
\label{fig1}       
\end{figure}
\begin{table*}
\caption{\label{data}Spherical beads of different materials as used in our experiments, where $\mu_k$ and $\mu_s$ are the coefficients of kinetic or static friction and $\epsilon$ is the coefficient of restitution}
\begin{tabular}{WCCCCCC}
\hline\noalign{\smallskip}
Material & Diameter (mm) & Mass (g) & Density (g/cm\textsuperscript{3}) & $\mu_k$ & $\mu_s$ & $\epsilon$  \\
\noalign{\smallskip}\hline\noalign{\smallskip}
Glass (soda-lime) & 3.0 $\pm$ 0.1& 0.035$\pm$ 0.001 & 2.5	& 0.126 $\pm$ 0.005	& 0.144 $\pm$ 0.008	& 0.90 $\pm$ 0.03 \\
Zirconium & 3.000 $\pm$ 0.002	& 0.085 $\pm$ 0.0005 & 4.49	& 0.124 $\pm$ 0.005 & 0.152 $\pm$ 0.007 & 0.76 $\pm$ 0.04 \\
Stainless Steel	& 3.171 $\pm$ 0.002 & 0.133 $\pm$ 0.0005 & 7.28	& 0.159 $\pm$ 0.008 & 0.20 $\pm$ 0.01 & 0.78 $\pm$ 0.04 \\
Steel Shot & 3.30 $\pm$ 0.04 & 0.150 $\pm$ 0.005 & 7.28	& 0.205 $\pm$ 0.006 & 0.22 $\pm$ 0.01	& 0.67 $\pm$ 0.05 \\
\noalign{\smallskip}\hline
\end{tabular}
\end{table*}

Our experiment involves adding individual beads to a preformed conical pile of the same beads on a circular base of fixed diameter (17.8 cm or about 60 bead diameters).  Each base had (with one exception described below) a monolayer of randomly placed beads glued to it that prevented the beads in the pile from rolling off the base.  The pile is initially formed by slowly pouring beads onto the base after which beads are added one at a time with the equilibrium mass of the pile captured by a LabVIEW data acquisition program. The mass of the pile is measured with a precision of 0.01~g, which is sufficient to resolve bead additions or the loss of one or more beads (as in an avalanche). The data acquisition program waits for the pile mass to stabilize before recording the mass and then dropping the next bead. Thus the system has the very slow driving rate needed for SOC.  The apparatus rests on a vibration isolated, optical table and is surrounded by a plexiglass box to minimize air drafts and acoustic noise.  In the first set of experiments, beads were dropped onto the apex of the pile from a height $h$ that can be adjusted from 0.6 cm to 11 cm and was measured by a cathetometer with a reproducibility of 1 mm.

Each data run consisted of dropping approximately 20 000 beads over a 54-hour period with multiple, reproducible runs to allow calculation of statistical error bars.  Fig.~\ref{fig1}(a) shows an example run of beads dropping onto the apex of the pile from a height of 1.5 cm.  Decreases in mass correspond to avalanches.  A narrow slice of data corresponding to 80 bead drops is shown enlarged in the inset; a pile-building period including a few one-bead avalanches and a three-bead avalanche can be seen at the left of the inset, followed by a large drop corresponding to a 30-bead avalanche, after which the pile again builds. An analysis routine is used to determine avalanche size and to count the number of avalanches of a given size and calculate the probability.  (Note that one-bead avalanches are not included in our calculations of the distribution $P(s)$ because they may be caused by the dropped bead falling immediately off the pile, by the dropped bead knocking a single other bead off the pile, or by a jam in the bead dropper so that no bead was dropped.) The typical size of our 3D bead pile was around 15 000 beads.  Details of the equipment and procedure were described previously \cite{CostelloPRE} where we reported the results of a similar experiment using 3.0 mm glass spheres.
We have repeated the experiment using beads of different densities (steel\footnote{New England Miniature Ball Company} and zirconium\footnote{K-Style Advanced Ceramics, Qingdao, People's Republic of China}) but the same nominal diameter of 3 mm.  Table~\ref{data} gives the relevant details for each of the beads we have used.  We include our measured values for the coefficients of kinetic  and static friction, $\mu_k$ and $\mu_s$, as well as the coefficient of restitution $\epsilon$, which measures the inelasticity of the material when involved in a collision \cite{LunAM}. 

As mentioned above, we take the change in mass of the pile as proportional to avalanche size.  While experiments on rice \cite{FretteNature} observed internal avalanches which did not change the pile mass, spherical beads have a different dynamical behavior.  When studying spherical glass beads on an inclined surface, Daerr and Douady \cite{DaerrNature} found that avalanches on the surface form a front that moves at an approximately constant velocity and incorporates the first couple of layers on the surface.  Borzsonyi et al. \cite{BorzsonyiPRL} found a different type of flow for irregular material than for spherical glass beads; the latter had fronts with a continuous structure traveling with a front speed that was faster than the moving beads.  Both studies involved beads on an incline past the critical angle of repose and, therefore, indicate that avalanches in spherical beads, once formed, continue moving down the surface of the pile and off the edge.

We have here investigated three experimental variations where identical beads were added to a pile to perturb its equilibrium. The first variation probed the effect of the bead's density on the avalanche size distribution as the drop height $h$ onto the apex of the pile was varied. In the second variation, we explored the effect of adding beads across the surface of the pile.  Finally, we note the emergence of structure within the pile when the base is a 2D hexagonal close-pack instead of random. 

\subsection{Apex dropping}
\label{apex_exp}
Individual beads were dropped onto the apex of the conical pile from an adjustable height $h$ above the apex.  We used both stainless steel  and then zirconium beads and varied the drop height from 0.6 cm to 11 cm and from 1.5 cm to 10 cm respectively.  Prior to use, the beads were cleaned using Micro-Soap and distilled water and dried thoroughly.

\subsection{Pseudo-random dropping}
\label{random_exp}

Another experimental variation explored dropping zirconium beads, not just onto the apex, but at varied locations across the pile.  We modified our bead dropper by adding a randomizer that consisted of four horizontal diamond-pattern metal screens vertically separated by 3.2 cm.  Zirconium beads dropped into this randomizer bounced chaotically before passing through the diamond-shaped openings to the next screen and finally emerging from the bottom at a semi-random location and with no vertical velocity.  We measured the distribution of beads coming out of the randomizer and found a broad Gaussian distribution centered above the bead pile's apex.   The diameter of the drop area was varied from 3.2 cm up to a maximum of 15.2 cm, which is slightly less than the overall pile diameter of 17.8 cm.  The bottom of the randomizer was kept at a constant height of 2 cm above the apex of the pile. Beads falling onto the edge of the pile fell a slightly longer distance, up to a maximum of 5.5 cm at the edge of the largest drop area.

\subsection{Emergence of structure}
\label{structure}

For one set of measurements with very uniform stainless steel beads, the base layer of beads was carefully glued in a hexagonal close-pack structure.  Stainless steel beads were then poured onto the pile forming the typical random packing that resulted in an angle of repose around 22\degree.  However, after many thousands of beads had been dropped onto the apex of the pile, we noted that a hexagonal structure emerged within the pile.  A close-pack triangular structure had previously been reported in 2D piles \cite{AltshulerPRL} but not in 3D piles.  In our piles, the emergent structure appeared at the bottom edge of the pile and worked up into a ÒcliffÓ at 52\degree or 66\degree that corresponded on alternating sides to face-centered cubic (fcc) and hexagonal close packing (hcp) structures.  Each ÒcliffÓ became 3-7 bead layers high and all six sides associated with the hexagonal bead pattern on the base could be observed with some random packing above this structure.  

Interestingly, the avalanche size distribution did not change as a result and we observed the avalanches to occur most often along the six boundaries between the ÒcliffsÓ, which themselves were quite robust once they formed.  This is in contrast to an experiment by Lorincz and Wijngaarden \cite{LorinczPRE} who built a pile in a square box with one open side and found quasi-periodic large avalanches when the pile came in contact with the three constraining sides.  Their pile then avalanched along one direction (toward the open end) and became essentially a 1D flow while ours remained a 2D flow as the avalanches typically occurred along the six edges on the surface of the pile.  We did not observe quasi-periodic large avalanches in our experiment and the avalanche size distributions when ÒcliffsÓ formed are consistent with, and included among, the stainless steel data taken using a randomly packed base shown in Fig.~\ref{fig2}.

\section{Results}
\label{results}

\subsection{Apex dropping}
\label{apex}

\begin{figure}
 \includegraphics{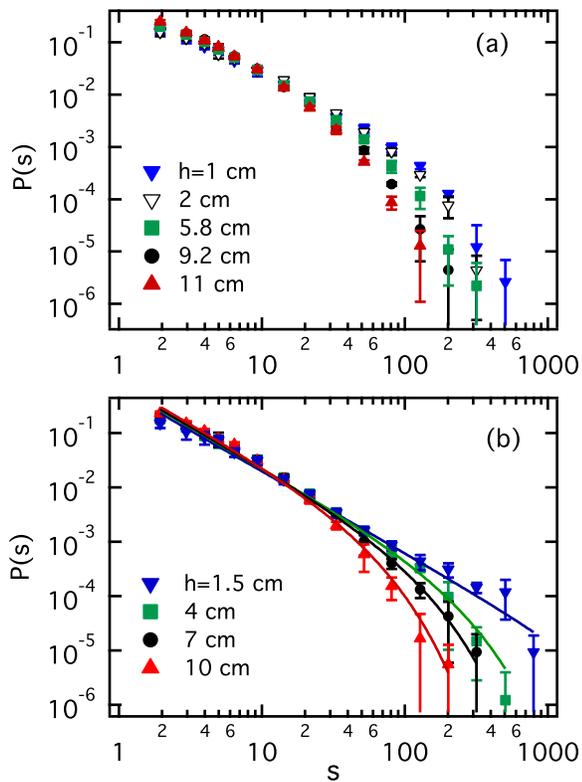}
\caption{The probability $P(s)$ of avalanches of size $s$  for (a) stainless steel beads and (b) zirconium beads.  The probability changes from a pure power law for small drop heights $h$ to being described by Eq. (\ref{powerlawRollOff}) for larger drop heights.  Lines indicate the result of weighted fits to Eq.~(\ref{powerlawRollOff}) with $\tau = 1.5$; fit lines are not shown in (a) for  clarity. Error bars are one standard deviation estimates }
\label{fig2}       
\end{figure}

For small drop heights with either bead material, the distribution of avalanche sizes are described by the pure power law of Eq. (\ref{powerlaw}) with an exponent $\tau$ of 1.5, which was the value found using glass beads \cite{CostelloPRE}.  For these small drop heights, the log-log plots of the distributions were approximately linear for two decades in avalanche size, more than similar experiments but less than the three or more decades needed to determine an exponent precisely.  As $h$ increased, the probability $P(s)$ decreased for larger avalanches.  This is shown in Fig.~\ref{fig2}(a) for stainless steel and \ref{fig2}(b) for zirconium.  A weighted fit to Eq. (\ref{powerlawRollOff}) can nicely describe the avalanche size distribution for beads dropped from larger heights as shown in Fig.~\ref{fig2}(b).  As with glass beads, the cutoff parameter  decreases as $h$ increases since $s_o$  roughly corresponds to the value of $s$ where the probability breaks from a pure power law. Remarkably, within our resolution,  $1/s_o$ is proportional only to $h$ and does \emph{not} depend on the mass of the bead (which differed by almost a factor of four), the coefficient of friction, or the coefficient of restitution.  

\subsection{Pseudo-random dropping}
\label{random}

The shape of the pile and the kind of avalanching was remarkably different as a result of using the randomizer.  Our pile was initially conical from simply pouring beads onto the base to form the initial pile, but as beads were added by the randomizer the pile became flatter and had many small avalanches and far fewer larger ones. The raw mass versus time plot for one of these runs is shown in Fig.~\ref{fig1}(b) where the small variation in pile mass and the lack of clear building periods reflect the many small avalanches that occurred.  The change in pile-building behavior can be seen in more detail in the inset of Fig.~\ref{fig1}(b) (set at the same scale as in Fig.~\ref{fig1}(a)); in addition to the few-bead avalanches visible, a substantial number of one-bead avalanches prevent the pile from building smoothly. 

\begin{figure}
	  \includegraphics{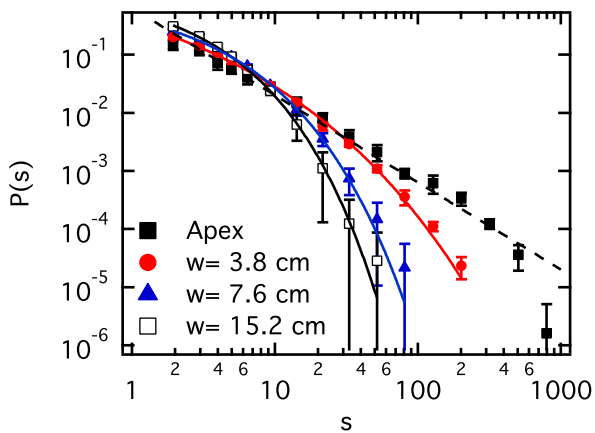}
	\caption{Zirconium beads dropped from a height of 2 cm (as measured from the apex of the pile) but distributed over a width $w$ of the pile.  Apex dropping data and dashed line corresponding to a power law with $\tau = 1.5$ are shown for comparison.  Other lines are guides for the eye}
\label{fig3}       
\end{figure}

The lack of large avalanches is apparent in the plot of the avalanche size distributions shown in Fig.~\ref{fig3}.  The dramatic roll-off of the probability was initially surprising so we varied the diameter $w$ of the randomizer by factors of two from 15.2 to 3.8 cm to look for systematic effects.  As expected, as the diameter of the region over which the beads were dropped approached apex dropping, the avalanche size distribution approached the distribution seen for apex dropping from the same height. The avalanche size distribution for the smallest drop area, $w = 3.8$ cm, was well described by Eq. (\ref{powerlawRollOff}), but the distributions for larger $w$ values were not well fit by this  modified power law.  

The change in pile-building behavior we observed is somewhat similar to that seen in a previous granular simulation which briefly investigated the effects of adding particles across the surface of a pile \cite{DendyPRE}.  Overall the simulation did not produce a realistic distribution of avalanche sizes, but the authors did note that random sprinkling caused pile growth to be interrupted by small avalanches while central (apex) ÒfuelingÓ made the pile grow uninterrupted.

\section{Discussion}
\label{discussion}

Our bead piles reach a critical state when the added beads arrive with little velocity and settle onto the pile, building it to a locally critical state which can then avalanche with the addition of a single bead that subsequently carries beads off the surface of the pile.  For beads dropped from small heights, we saw the mass of the pile generally increasing along with many small avalanches as well as the occasional large avalanche as can be seen in Fig.~\ref{fig1}(a).  The resulting avalanche size distribution was well described by a power-law.  However, when beads were added to the apex of the pile with larger velocities, some beads that were locally sub-critical on the pile were dislodged resulting in more frequent small and intermediate sized avalanches.  These prevented the pile from building enough for large avalanches to occur.  The system was kept sub-critical and the avalanche size distribution rolls off for large avalanches as seen in Fig.~\ref{fig2}.  

\begin{figure}
 	 \includegraphics{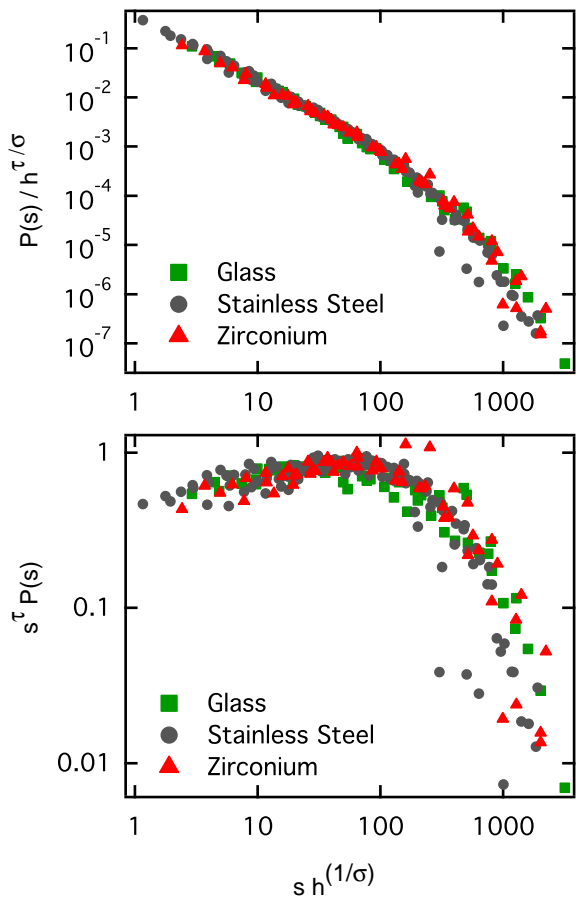}
	\caption{The avalanche size distribution data for all types of beads falling from all drop heights $h$ onto the apex of a pile collapses onto a common curve when scaled using Eq. (\ref{powerlawScaling}) as described in the text. A plot of $P(s) h^{-\tau/\sigma}$ versus $sh^{1/\sigma}$ is shown in (a); $P(s) s^\tau$ versus $sh^{1/\sigma}$ is plotted in (b).  The values of $\tau = 1.5$ and $\sigma = 1$ were found to give the best collapse in both plots}
	\label{fig4}       
\end{figure}

\paragraph*{Scaling Analysis}A better way of representing the behavior of the granular pile when dropping beads onto the apex is by scaling the distribution data as described in Sect.~\ref{intro}.  Specifically, if $h$ is a tuning parameter, then Eq.~(\ref{collapse1}) and Eq.~(\ref{collapse2}) can be used to provide a collapse of all the distribution data at all drop heights and all bead types onto a common curve.  This is shown in Fig.~\ref{fig4} using $\tau = 1.5$ and $\sigma = 1$, which best collapsed the data onto the two common curves shown in (a) and (b).  That the avalanche size distribution for beads of very different densities, coefficients of restitution, and coefficients of friction would collapse onto a common curve using such simple scaling indicates that the instability structure of the conical pile in 3D is effectively independent of the beads used.  The scaling parameter $h$ allows the system to be brought close to the critical point when $h$ is small but smoothly taken away from critical as $h$ increases.

\begin{figure}
  \includegraphics{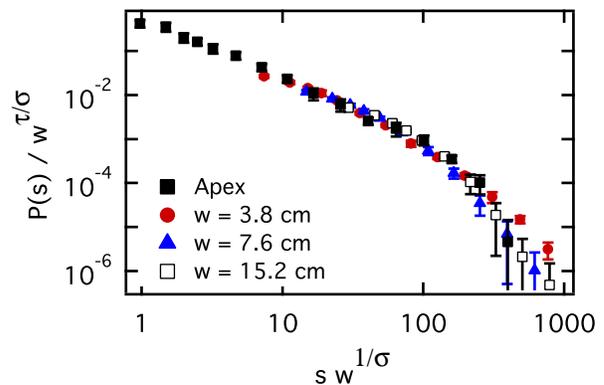}
\caption{The avalanche size distributions for beads falling across the surface of the pile shown in Fig.~\ref{fig3} can be scaled according to Eq. (\ref{collapse1}) (using $w$ in place of $h$) so all the data fall onto a common curve.  The exponent values are the same as in Fig.~\ref{fig4}.  The apex data were scaled using $w=0.5$ cm}
\label{fig5}       
\end{figure}

A similar effect is seen when zirconium beads were dropped over a region of width $w$ onto the bead pile from a fixed height.  The same kind of scaling described above (when $h$ was used as a parameter in Eq. (\ref{powerlawScaling})) also works when the tuning parameter becomes $w$.   The avalanche size distributions shown in Fig.~\ref{fig3} scale onto a common curve as seen in Fig.~\ref{fig5} using this scaling approach when $\tau = 1.5$ and $\sigma = 1$.  This indicates that varying the region over which beads are dropped acts as another tuning parameter for the system in a way that is analogous, but not equivalent, to varying the height when dropping onto the apex. 

\begin{figure}
  \includegraphics{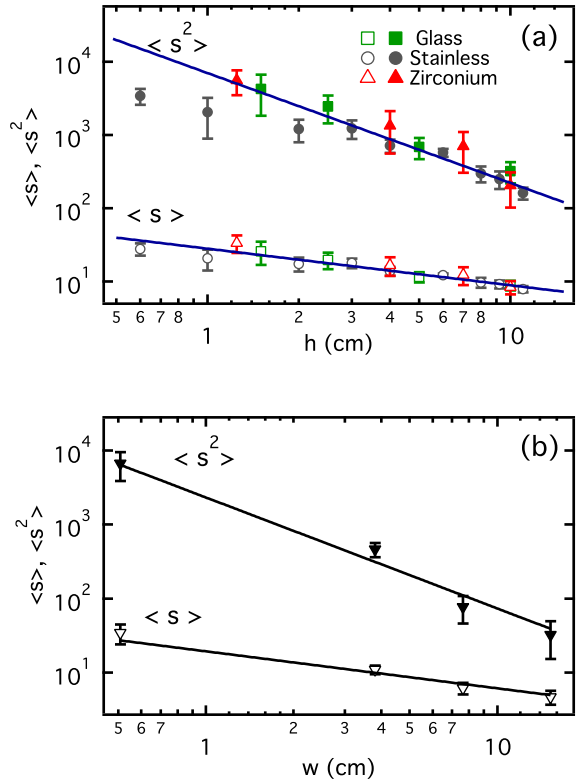}
\caption{The moments  $\langle s \rangle$ and $\langle s^2 \rangle$  of the avalanche size distribution as a function of (a) drop height $h$ above the apex of the pile for glass, stainless steel and zirconium beads and (b) width $w$ across the pile for zirconium beads.  The lines show an $h^{-1/2}$ (or $w^{-1/2}$) dependence for $\langle s \rangle$  and an $h^{-3/2}$ (or $w^{-3/2}$) dependence for  $\langle s^2 \rangle$.  The power-law behaviors confirm scaling in this system }
\label{fig6}       
\end{figure}

We can further test the robustness of the scaling when $h$ or $w$ act as a tuning parameter as well as testing the value of the exponents $\tau$ and $\sigma$ by calculating the two moments $\langle s \rangle$ and $\langle s^2 \rangle$  of the avalanche size distribution as detailed in the literature \cite{LiuEPL,SethnaNature}.  If scaling holds as discussed in Eq. (\ref{moments}) then the moments of the distribution $\langle s \rangle$ and $\langle s^2 \rangle$ should vary as $h^{\left(\tau-2\right)/\sigma}$ and $h^{\left(\tau-3\right)/\sigma}$
 respectively, which, for $\tau = 1.5$ and $\sigma = 1.0$, correspond to $h^{-1/2}$ and $h^{-3/2}$ (or to $w^{-1/2}$ and $w^{-3/2}$).   Using the discrete values from our measured probabilities, we can calculate the moments from each of the data runs presented in Figs.~\ref{fig2} and \ref{fig3}; these values are plotted in Fig.~\ref{fig6}(a) against $h$ and Fig.~\ref{fig6}(b) against $w$.  In both cases, the data show very good consistency with a power-law using a power of ${\sfrac{-1}{2}}$ for  $\langle s \rangle$ and ${\sfrac{-3}{2}}$ for $\langle s^2 \rangle$  as shown by the solid lines.  The only deviation appears in the $\langle s^2 \rangle$  values for the three smallest drop heights with one type (stainless steel) of bead; if those three points are excluded from a weighted fit to the other $\langle s^2 \rangle$ points in (a) then we obtain a power of $1.55\pm 0.13$.  A weighted fit to the $\langle s \rangle$ points from (a) yields a power of $-0.45\pm 0.05$.  Inverting these two results for $\tau$ and $\sigma$, we find $\tau = 1.6 \pm 0.1$ and $\sigma = 0.91 \pm 0.15$. This calculation is independent from the collapse technique of Eq. (\ref{collapse1}) and Eq. (\ref{collapse2}) but yields values of exponents consistent with our finding of $\tau$ = 1.5 and $\sigma$ = 1 from Fig.~\ref{fig4}.

\section{Conclusion}
\label{conclusion}

With our randomly packed, 3D conical bead piles of spherical beads on a base of approximately 60 bead diameters, we find the avalanche size distributions to indicate critical point behavior in this non-equilibrium system.  At small drop heights onto the apex of the pile, the distributions for glass, zirconium, or stainless steel beads are described by a simple power-law with exponent $\tau = 1.5$.  We find the drop height $h$ and the diameter of the drop area $w$ to be two independent tuning parameters that take the system away from the critical point.  Despite differences in the inertia, surface properties, and elasticity of the beads, all the distributions collapse onto a common curve when scaled appropriately by the tuning parameter and using $\tau = 1.5$ and $\sigma = 1$.  The moments of those distributions show a power-law behavior in the tuning parameters that further show the scaling behavior of the system and give a value for the exponents $\tau = 1.6 \pm 0.1$ and $\sigma = 0.91 \pm 0.15$. This value for $\tau$ is surprisingly close to the mean-field value predicted for systems of four or more dimensions.

\begin{acknowledgments}
We are grateful to Amy Braun and Karin Dahmen for discussions and for bringing certain papers to our attention.  Tuan Nguyen helped with the experiment.  The research was supported by The College of Wooster, Research Corporation, and NSF REU grant DMR-0649112.
\end{acknowledgments}

\end{document}